\documentclass[prb,twocolumn,amsmath,amssymb,floatfix]{revtex4}
\usepackage{graphicx}

\newcommand{\be}{\begin{eqnarray}}
\newcommand{\ee}{\end{eqnarray}}
\newcommand{\bfr}{{\bf r}}

\newcommand{\hb}{\hat{b}}
\newcommand{\ha}{\hat{a}}
\newcommand{\hn}{\hat{n}}
\newcommand{\hS}{\hat{S}}
\newcommand{\wbe}{\begin{widetext}}
\newcommand{\wee}{\end{widetext}}
\newcommand{\oncite}{\onlinecite}

\begin{document}
\draft

\title{Quantum phase transitions 
of polar molecules in bilayer systems}
\author{Daw-Wei Wang}

\address{Physics Department, National Tsing-Hua University, Hsinchu, 
Taiwan, ROC
}

\date{\today}

\begin{abstract}
We investigate the quantum phase transitions of bosonic polar molecules
in a two-dimensional double layer system. We show that an interlayer
bound state of dipoles (dimers) can be formed when 
the dipole strength is above a critical value, leading to
a zero energy resonance in the interlayer $s$-wave scattering channel. 
In the positive detuning side of the resonance, the strong {\it repulsive}
interlayer pseudo-potential can drive the system into 
a maximally entangled state, 
where the wavefunction is a superposition of
two states that have all molecules in one of the two layers
and none in the other. We critically discuss how the zero-energy resonance,
dimer states and the maximally entangled state can be
measured in time-of-flight experiments.
\end{abstract}


\maketitle
{\it Introduction:}
Systems of ultracold atoms have become one of
the most promising systems to observe strong correlation 
effects in many-body physics. 
Recent progress in the trapping and cooling of chromium atoms
[\onlinecite{Cr}] and polar molecules [\onlinecite{cool_molecule}] further
opens new directions for investigating quantum many-body states
resulting from the anisotropic dipole-dipole 
interaction [\oncite{dipole_exotic}].
The long range nature of dipole interaction also makes
it possible to study physics in spatially separated 
multi-component systems, which have been extensively studied
in several important subfields of solid state physics: for example, 
condensation of
excitons in bilayer quantum well system [\oncite{exciton}],
interlayer ferromagnetism in bilayer quantum Hall systems [\oncite{QH}] and 
Coulomb drag in coupled quantum wires [\oncite{drag}] etc.
Therefore it is interesting to investigate what new physics one
may expect in the similar systems of polar molecules.
Recent example is the proposed chaining phenomena for
molecules in a stack of 2D traps [\oncite{wang}], which
resembles particle aggregation in colloidal fluids [\oncite{x}]. 

\begin{figure}
\includegraphics[width=8cm]{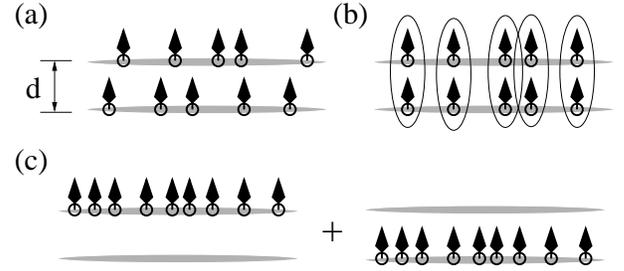}
\caption{
Three many-body states we consider in this paper: 
(a) coupled superfluid state, (b) superfluid of
dimers, and (c) maximally entangled state.
}
\label{states}
\end{figure}

In this paper we investigate the quantum phase transitions of 
cold polar molecules trapped in a 2D double well potential
[\oncite{2D_layer}].
The electric dipole moment ($D$) is aligned perpendicular to the
layer ($x-y$) plane by a DC electric field (see Fig. \ref{states}(a))
so that the system properties is controlled by a dimensionless
dipole strength, $U_0\equiv m D^2/\hbar^2 d$, with $m$ being the 
molecule mass and $d$ being the layer separation. We find three
phases that can be observed in three different regimes of $U_0$:
For weak dipole strength, $U_0\ll 1$, the ground state is just
a coupled superfluid (Fig. \ref{states}(a)).
When $U_0$ is increased to be above a critical value, $U_0^\ast\sim 0.71$,
molecules in different layers can form interlayer bound states, 
driving the system to be a superfluid of dimers (Fig. \ref{states}(b)).
Finally, if the molecules are cooled in the large $U_0$ regime and
the dipole moment is reduced toward $U_0^\ast$ adiabatically {\it from above}, 
we demonstrate that the {\it repulsive} interlayer 
pseudo-potential can drive the
system to a maximally entangled 
state as $U_0\to U_0^{\ast\ast}\sim 1.4$, breaking a global
$U(1)$ symmetry via a second order transition.
Such maximally entangled (ME) state 
is a superposition of two macroscopical 
states (or called GHZ state [\oncite{ME}])
that have all molecules in one layer and none in the other 
(Fig. \ref{states}(c)) [\oncite{cat_coupling}], and therefore
will not have any interference 
pattern even in a single shot time-of-flight measurement.

{\it Pseudo-potential:}
We start from the low energy scattering properties between two
molecules via dipole interaction.
In the strong confinement regime, we can first assume 
only the lowest subband of each layer
is occupied and no single particle tunneling between them.
The 2D Schr\"{o}dinger equation in the relative coordinate
can then be written to be
\be
-\frac{\hbar^2}{m}\left(\partial_x^2+\partial_y^2\right)\phi(\bfr)
+V_{0/1}(\bfr)\phi(\bfr)
&=&E\phi(\bfr)
\ee
where $V_{0/1}(\bfr)\equiv\int dz_1dz_2 
\left|\varphi_0(z_1-d/2)\right|^2
\left|\varphi_0(z_2\mp d/2)\right|^2$
$V_d(\bfr,z_1-z_2)$
is the bare interaction for the two molecules
in the same/different layers.
$V_{d}(\bfr,z)=D^2\left(\bfr^2-2z^2\right)/(\bfr^2+z^2)^{5/2}$
is the dipole interaction with $\bfr$ being the relative 
coordinate in the $x-y$ plane. $\varphi_0(z)$ is the lowest
confined wavefunction and can be approximated by a Gaussian 
wavefunction of width $W$ ($W\ll d$). When finite interlayer 
tunneling ($t$) is considered, one has to diagonalize the full two-particle
two-layer Hamiltonian. Its effect to Eq. (1) can be
shown to be the order of $t^2/(D^2/W^3)$, and
hence negligible in the strong confinement regime as we considered here.

Now, using the standard scattering theory [\oncite{yang}], 
we can derive the following 2D pseudo-potential:
\be
{\cal V}^{(0)/(1)}_{\rm ps}(\bfr)
&=&-\frac{4\hbar^2}{m}\tan\delta^{(0)/(1)}_0(k)\cdot
\delta(\bfr),
\label{pseudo-V}
\ee
which reproduces the same $s$-wave phase shift ($\delta^{(0)/(1)}_0(k)$) as
the bare interaction, $V_{0/1}(\bfr)$, in large distance.
$k$ is the magnitude of the incoming relative momentum. We can show
that contributions from higher angular momentum channels can be
safely neglected since the typical length scale of 
dipole interaction, $mD^2/\hbar^2$($\sim 1.5 \mu$m
for $D\sim 1$ Debye and $m\sim 100$ a.m.u.) is 
much smaller than the typical condensate size.

\begin{figure}
\includegraphics[width=8.5cm]{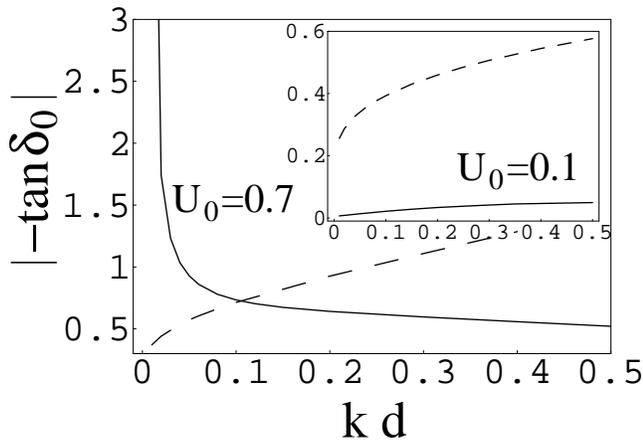}
\caption{
Intralayer (dashed lines) and interlayer (solid lines) scattering
phase shift in the $s$-wave channel as a function
of momentum $kd$ near resonance ($U_0=0.7$). 
The layer width, $W=0.1 d$, is set
much smaller than the interlayer distance $d$. 
Note that phase shift ($\delta_0$) of the interlayer scattering
is positive, and its sign is reversed for the
convenience of comparison with the intralayer results.
Inset: Results for $U_0=0.1$.
}
\label{U0_kd}
\end{figure}

\begin{figure}
\includegraphics[width=8.5cm]{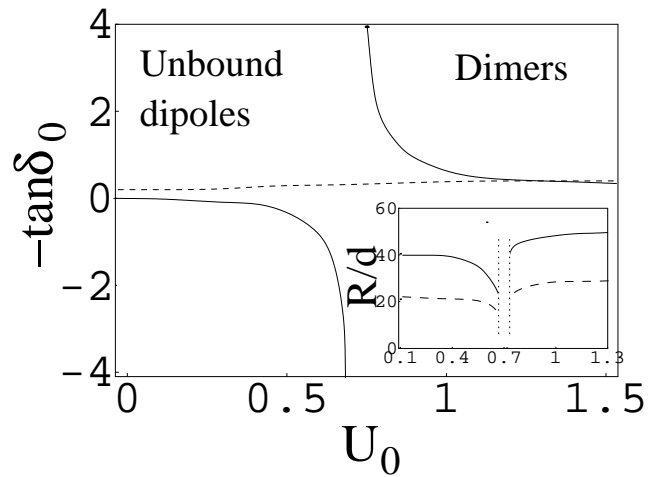}
\caption{
Zero-energy resonance of the interlayer(solid line) scattering phase shift 
as a function of $U_0$.
Dashed line is for the {\it intra}layer scattering.
Inset: Calculated condensate radius as a function of $U_0$.
Solid and dashed lines are for harmonic oscillator length, 
$a_{\rm ho}\equiv\sqrt{\hbar/m\omega_\|}=10 d$,
and $5d$ respectively. Number of molecules in each
layer is $N=10^5$ and other parameters are the same as 
in Fig. \ref{U0_kd}. Our meanfield
treatment of Eq. (\ref{E_MF}) may fails near resonance 
and therefore we eliminate the data between the two dotted lines.
}
\label{phase_shift_U0}
\end{figure}

{\it Zero energy resonance:}
In Fig. \ref{U0_kd}, we show the calculated $s$-wave
phase shift as a function of $kd$ for $U_0=0.7$ and 
$U_0=0.1$ (inset).  When the dipole strength is
weak(inset), the phase shift of the interlayer scattering is always
much smaller than that of intralayer one as expected, but
it becomes much larger when $U_0$ is larger.
In Fig. \ref{phase_shift_U0}, we show the numerically 
calculated low energy ($kd\to 0$) $s$-wave phase shift
as a function of $U_0$, and find a resonance at $U_0=U_0^\ast\sim 0.71$.
Similarly to the Feshbach resonance in typical 
cold atom systems [\oncite{FR}], this zero energy 
resonance is due to the formation of
an interlayer bound state (dimer). Interaction between two dimers can 
be obtained by integrating out the dimer wavefunction (not shown here).
The system ground state for $U_0>U_0^\ast$
is a superfluid of dimers with a finite binding energy
(about $0.1\times\hbar^2/md^2$ at $U_0=1$), 
but the actual phase transition 
position might be shifted from $U_0^\ast$ due to the interaction between
dimers. We also note that the quantum phase transition from a coupled
superfluid to dimer superfluid near $U_0^\ast$ is belong to the Ising
type transition, because the dimer superfluid phase just breaks 
a $U(1)/Z_2$ symmetry, similar to bosnoic systems near Feshbach
resonance as discussed in Ref. [\oncite{sachdev}].
We point that these result cannot be reproduced even qualitatively within the
Born approximation in the literature [\oncite{dipole_simple}], which 
is valid only when the dipole strength is very weak ($U_0\ll 1$).

{\it Condensate size near resonance:}
It is interesting to study how the condensate size is changed when
the dipole strength $U_0$ is tuned across the resonance point.
Using a Gaussian variational wavefunction [\oncite{Yu}], 
$\Psi_\pm(\bfr)=\frac{\sqrt{N}}{\sqrt{\pi} R}\,e^{-|\bfr|^2/2R^2}
\varphi_0(z\mp d/2)$, for the condensate wavefunction in 
the upper(+) and the lower(-) layers,
the radius $R$ in the negative detuning side 
is then obtained by minimizing the following meanfield energy:
\be
\frac{E}{N}
&=&-t+\frac{\hbar^2}{mR^2}+m\omega_\|^2 R^2
+\frac{N\hbar^2}{8\pi m R^2}\sum_{i=0,1}A_0^{(i)}(R).
\label{E_MF}
\ee
Here $\omega_\|$ is the in-plane trapping
frequency, and $A^{(0)/(1)}_0(R)\equiv -\int_0^\infty x
\tan\delta_0^{(0)/(1)}(x/R)\,e^{-x^2/4}$ are the dimensionless
interaction energies. $t$ is single particle tunneling amplitude and here
contributes to a constant only within meanfield approximation.
We can also apply a similar method to describe the condensate size
of dimers in the positive detuning side ($U_0>U_0^\ast$). 
When $U_0$ is well above $U_0^\ast$ (i.e. large binding energy of dimers), 
the low energy scattering does not break a dimer and the phase sift
can be calculated from the interaction between dimers after taking into 
account their bound state wavefunction [\oncite{unpublished}].
In the inset of Fig. \ref{phase_shift_U0}, we show the calculated
condensate radius as a function of $U_0$.
In the negative detuning side,  
the condensate size decreases gradually as $U_0$ approaching
$U_0^\ast$ from below due to the increasing attractive interlayer 
pseudo-potential (main plot). On the other (positive detuning) side, 
the size of the dimer condensate grows rapidly due to
the repulsive interaction between dimers. Although the
meanfield calculation may not be reliable when very close to 
the resonance regime due to the strong momentum dependence of 
interlayer pseudo-potential, it is reasonable to expect that
the sharp shrinking of condensate size 
near $U_0^\ast$ is still qualitatively true.
Therefore measuring the dramatical change of condensate size near 
resonance can provide a clear evidence of zero energy resonance as well
as the dimer state in the bilayer system.

{\it Maximally entangled state:}
In the previous discussion we concentrated on the situation
where the dipole strength $U_0$ is initially
small and adiabatically increased to be above $U_0^\ast$.
However, in a realistic experiment, the electric dipole
moment can be so strong that molecules are cooled directly
in the large $U_0$ regime with very small transition rate
to the dimer state. 
It is therefore interesting to study how the many-body {\it metastable} state
is changed when the dipole strength $U_0$
is adiabatically tuned toward the critical value ($U_0^\ast$) from {\it above}.
From Fig. \ref{phase_shift_U0} one can see that there are two 
regions of interest in this positive detuning side:
one is for $U_0>U_0^{\ast\ast}\sim 1.4$ where the
effective interlayer interaction is repulsive but still smaller 
than the intralayer interaction in the long wavelength limit, and
the other is for $U_0^\ast<U_0<U_0^{\ast\ast}$
where the interlayer pseudo-potential is larger than the
intralayer one. Since both inter- and intra-layer 
interactions are repulsive for $U_0>U_0^\ast$, 
hereafter we may neglect the in-plane trapping potential
and consider a homogeneous system for simplicity. Assuming all dipoles 
are in the zero momentum state at zero temperature,
we can write the following effective Hamiltonian
\be
H&=&-t\left(\ha_0^\dagger \hb_0^{}+\hb_0^\dagger \ha_0^{}\right)
+\frac{g_0}{2N}\left[\hn_a^2+\hn_b^2\right]
+\frac{g_1}{N}\hn_a\hn_b
\nonumber\\
&=&\frac{g_0}{2N}(2N)^2
-t\left(\ha_0^\dagger \hb_0^{}+\hb_0^\dagger \ha_0^{}\right)
+\frac{\Delta g}{N} \hn_a\hn_b,
\label{H_cat}
\ee
where $\ha^\dagger_0$($\hb^\dagger_0$) are boson creation operators 
in the upper/lower layer at $k=0$ with $\hn_a$ and
$\hn_b$ being their number operators.
$g_{i}\equiv\frac{-4\hbar^2N}{m\Omega}\tan\delta_0^{(i)}(k\to 0)$ 
is the intra-($i=0$) or inter-($i=1$) meanfield energy with $\Omega$
being the condensate area. $\Delta g\equiv g_1-g_0<0$ for 
$U_0>U_0^{\ast\ast}$ and $\Delta g>0$ for $U_0^\ast<U_0<U_0^{\ast\ast}$.
In the second line of Eq. (\ref{H_cat}) 
we have used $\hn_a+\hn_b=2N$ as the conserved total
number of molecules. 
We note that although Eq. (\ref{H_cat}) looks similar to the two-site 
Bose-Hubbard model with inter-site interaction, the physics described
by Eq. (4) is different from the well-known 
superfluid to Mott-insulator phase transition [\oncite{SF-MI}].
For example, the lowest excitation state
of our system is always the in-plane gapless phonon mode and therefore
no charge gap or commensurate filling are expected 
even when $t$ is reduced to zero.

Before showing the calculation results, it is instructive
to discuss the analytic solutions in three different limits: Firstly,
in the limit of $t\ll |\Delta g|$ and $\Delta g<0$, 
the ground state wavefunction is very close to a Fock state:
$|\Psi_{\rm Fock}\rangle=\frac{1}{N!}
\ha_0^\dagger{}^N \hb_0^\dagger{}^N|0\rangle$, with almost zero
interlayer phase correlation (or the phase stiffness is very weak).
Secondly, for $\Delta g=0$ but with finite tunneling, the ground state
is a condensed symmetric coherent state, $|\Psi_{\rm Sym}\rangle=
\frac{1}{2^N\sqrt{(2N)!}}
\left(\ha^\dagger_0+\hb^\dagger_0\right)^{2N}
|0\rangle$. The two 
condensate are now phase locked by the single particle tunneling so that
there will be a true phase correlation, which can be measured in
series of time-of-flight experiments.
Finally, in the limit of $\Delta g \gg t >0$, the total energy 
is minimized by $\langle \hn_a\hn_b\rangle=0$,
i.e. all dipoles are in one of the two layers and none
in the other. The most general ground state wavefunction is a 
superposition of two macroscopic states,
$|\Psi_{\rm ME}\rangle=\frac{1}{\sqrt{(2N)!}}\left(\cos\xi
a^\dagger_0{}^{2N}+\sin\xi e^{i\chi} b^\dagger_0{}^{2N}\right)|0\rangle$,
with tilted angle $\xi$ and phase $\chi$ being arbitrary. 
Such state is also known as a kind of Greenberg-Horne-Zeilinger state
[\oncite{ME}], which maximizes the entanglement in many measures.
Note that the
maximally entangled (ME) state we consider here is spontaneously 
generated as an exact eigenstate of the system, stabilized by the 
many-body effects.

\begin{figure}
\includegraphics[width=8cm]{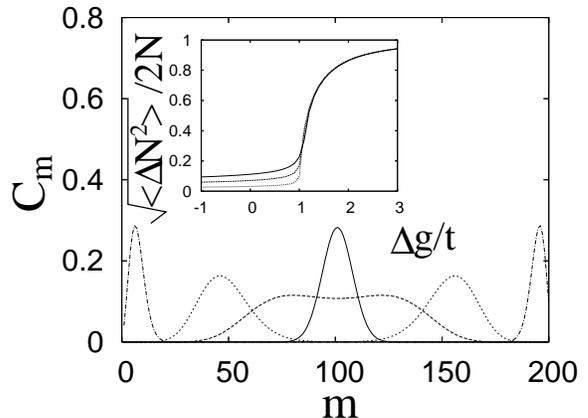}
\caption{
Many-body wavefunction, $C_m$, for
total number of $2N=200$ dipoles.
Solid, dashed, dotted, and dash-dotted lines are for 
$\Delta g/Nt=-3,1.05,1.2$, and 3 respectively. 
Inset: particle number variation
as a function of $\Delta g/t$. solid, dashed, and dotted lines are 
for dipole number $N=40$, 100, and 500 in each layer.
}
\label{fig_cat}
\end{figure}
The Hamiltonian of Eq. (\ref{H_cat}) can be easily diagonalized
in the Fock states basis:
$|\phi_m\rangle\equiv\left(m!(2N-m)!\right)^{-1/2}
a^\dagger_0{}^m b^\dagger_0{}^{2N-m}|0\rangle$, where 
$m=0,1,\cdots,2N$ is the number of dipoles in the upper layer.
The ground state wavefunction can therefore be written
to be $|\Psi_G\rangle=\sum_{m=0}^{2N}C_m|\phi_m\rangle$ with $\{C_m\}$ being
the eigenvector associated with the lowest eigenenergy of Hamiltonian
matrix, $\langle\phi_m|H|\phi_n\rangle$. Similar approach can also be  
applied to systems of finite trapping potential.
In Fig. \ref{fig_cat} we show the exact numerical results
of the ground state wavefunction ($C_m$) for different values
of $\Delta g/t$.
One can see that for smaller $\Delta g/t$ (solid line), the 
wavefunction is peaked at $m=N$ with a finite distribution width to gain
tunneling energy. 
When $\Delta g/t$ is close to one, the single peak distribution 
becomes unstable and tends toward a
double-peak distribution. Increasing $\Delta g$ further (i.e. reducing
$U_0$ in the positive detuning side) drives the
distribution to peak near $m=0$ and $2N$, indicating an ME 
state as discussed above. In the inset 
of Fig. \ref{fig_cat}, we show the particle number variation of the
ground state, $\langle\Delta N^2\rangle\equiv
\langle \psi_G|(a_0^\dagger a_0^{}-b_0^\dagger b_0^{})^2
|\Psi_G\rangle$, as a function of $\Delta g/t$ for various numbers of
dipoles per layer ($N$). One can see that in the thermodynamic 
limit (i.e. keeping $\Delta g\propto N/\Omega$ fixed as $N\to\infty$), 
there is a sharp phase transition exactly at $\Delta g/t=1$, above which 
$\sqrt{\langle\Delta \hat{N}^2\rangle}/2N$ 
becomes finite and eventually saturates. 

To understand such a sharp phase transition from a coherent state
to the ME state, 
we can rewrite Eq. (\ref{H_cat}) as a spin
model (upto a constant) [\oncite{cat_coupling}]: 
$H=-2t\hS_x-\frac{\Delta g}{N}\hS_z^2$, where
$\hS_x\equiv\frac{1}{2}(a^\dagger_0b^{}_0+b^\dagger_0a^{}_0)$,
$\hS_y\equiv\frac{i}{2}(b^\dagger_0a^{}_0-a^\dagger_0b^{}_0)$,
and $\hS_z\equiv\frac{1}{2}(a^\dagger_0a^{}_0-b^\dagger_0b^{}_0)$.
The total spin is then given by 
$\hat{\bf S}^2=\frac{1}{4}(\hn_a+\hn_b)(\hn_a+\hn_b+2)$.
In the thermodynamic limit ($N\to \infty$), 
we can treat {\bf S} a classical spin: 
${\bf S}=N(\sin\theta\cos\phi,\sin\theta\sin\phi,\cos\theta)$ with $\theta$
and $\phi$ being the spin angles in 3D space.
Therefore the ground state is obtained by minimizing 
the energy $E(\theta,\phi)/N=-2t\sin\theta\cos\phi-\Delta g
\cos^2\theta$ with respect to $\theta$ and $\phi$.
Since $\phi$ must be zero to gain the
tunneling energy, we can expand $E(\theta,0)$ to the leading order of 
$\eta=\theta-\pi/2$ and obtain a Ginsberg-Landau type energy:
$E(\eta,0)/N=-2t+(t-\Delta g)\eta^2+\frac{1}{3}(\Delta g-t/4)\eta^4,$
which shows a clear second order phase transition at $\Delta g=t$. The 
variation of particle number (i.e. variation of $\hS_z$) scales as
$(\Delta g-t)^{1/2}$ near the transition point.

Before concluding, we remark on several experimental issues for observing the
maximally entangled state in the bilayer system. 
First, for a typical polar molecule
$U_0$ can be as large as 4-5 and can be easily reduced to zero by decreasing
the external DC electric field. Secondly we can show
that phase separation (i.e. dipoles accumulate inhomogeneously in layers) 
is unlikely to occur because it causes extra kinetic energy compare to the 
homogeneously entangled state.  
Thirdly, the three-body collisions induced transition to dimer states
can be strongly suppressed in the ME state, because most molecules now
are in one of the layers and very few molecules are in the other layer. 
As a result the ME state we proposed here 
should be a long-lived meta-stable state and hence
can be easily observed in experiments.
Finally, unlike the interference pattern of two independent 
condensates [\oncite{castin}],
the fringe contrast of such entangled state will disappear even in 
a {\it single-shot} time-of-flight measurement as $U_0$ is adiabatically
tuned to be lower than $U_0^{\ast\ast}$ from above.
This is because the maximally entangled state, $|\Psi_{MF}\rangle$, 
is a superposition of two macroscopic states,
and hence very fragile to collapse in any quantum measurement.
Therefore the disappearance of interference pattern in the positive detuning
side could be a direct experimental evidence of 
such maximally entangled state.

In summary, we demonstrate that loading polar molecules into a bilayer system
can result in several interesting new physics, including
zero energy resonance, interlayer bound states
(dimers), and a second order quantum phase transition toward a 
maximally entangled state.
These new phenomena should be easily observable using present experimental
techniques.

This work is supported by NSC (Taiwan).


\end{document}